\documentclass{aa}
\usepackage{natbib}
\usepackage{txfonts}
\bibpunct{(}{)}{;}{a}{}{,}%
\usepackage{psfrag}
\usepackage{amssymb}
\usepackage[dvips]{graphicx}
\newcommand{\pr}{\mathrm{Pr}}
\newcommand{\ray}{\mathrm{R}}
\newcommand{\re}{\mathrm{Re}}
\newcommand{\aspra}{A}
\renewcommand{\div}{\vec{\nabla}\!\cdot\!}
\newcommand{\grad}{\vec{\nabla}\!}

\title{Mesoscale flows in large aspect ratio simulations of\\ turbulent compressible convection}

\author{F. Rincon \and F. Ligni\`eres \and  M. Rieutord}
\institute{Laboratoire d'Astrophysique de Toulouse-Tarbes, UMR CNRS
  5572, Observatoire Midi-Pyr\'en\'ees, 14 avenue \'E. Belin, 
  31400 Toulouse, France}
\offprints{F. Rincon\\\email{rincon@ast.obs-mip.fr}}

\date{Received \today / Accepted }

\begin{document}

\abstract{We present the results of a very large aspect ratio ($\aspra=42.6$) 
  numerical simulation of fully compressible turbulent convection in a
  polytropic atmosphere, and focus on the properties of large-scale
  flows. Mesoscale patterns dominate the turbulent energy spectrum.
  We show that these structures, which had already been observed in
  Boussinesq  simulations by \cite{cattaneo01}, have a genuine
  convective origin and do not result directly from collective
  interactions of the smaller scales of the flow, even though
  their growth is strongly affected by nonlinear transfers. If this 
  result is relevant to the solar photosphere, it suggests that 
  the dominant convective mode below the Sun's surface may be at mesoscales. 

}
\authorrunning{F. Rincon et al.}
\titlerunning{Mesoscale flows in large aspect ratio simulations of turbulent compressible
  convection}

\maketitle

\section{\label{intro} Introduction}
The origin of solar photospheric flows on horizontal scales larger than 
granulation ($\ell\sim 1~000$ km) has been a puzzling problem for more than 
forty years, when supergranulation ($\ell\sim 30~000$ km) was
discovered by \cite{hart54} and later on confirmed by \cite{simon64}. 
Even though recent breakthroughs in the field of supergranulation imaging have been
made thanks to the emergence of local helioseismology techniques 
\citep{duvall2000} and the results of the MDI instrument 
\citep{hathaway2000}, its origin is still unclear.
The existence of an intermediate scale, mesogranulation
($\ell\sim8~000$ km), is also a matter of debate 
\citep{hathaway2000,shine2000,rieutord2000,lawrence2001}. 

Meso and supergranulation have long been believed to be due to 
Helium deep recombinations driving cell-like convection.
This view now appears to be out of date
\citep{rast03b}. Several numerical experiments of
convection \citep{cattaneo91,nordlund00} at moderate aspect ratio
($A$ is the ratio of the box width to its depth) have shown a
tendency of long-lived large-scale flows to form in depth. Using large
aspect ratio ($\aspra=20$) simulations of Boussinesq
convection, \cite{cattaneo01} have suggested that mesogranulation
may result from nonlinear interactions of granules  
\cite[see also][]{rast03b}.
A large scale instability \citep{gama94} of granules 
has also been proposed by \cite{rieutord2000} to explain
supergranulation. Local numerical simulations at $A=10$
\citep{rieutord02} did not confirm it. 
\cite{derosa02}, using spherical simulations,
have computed flows down to supergranular scales.
Actually, the emergence of the three distinct scales of granulation,
mesogranulation, and supergranulation in the surface layers, among the observed 
\textit{continuum} of scales, remains a fully open problem that still 
deserves much work.

In this Letter, we report new results on three-dimensional numerical 
simulations of fully compressible  turbulent convection in 
a rectangular box with very large aspect ratio $\aspra=42.6$. This
configuration allows us to study accurately the turbulent dynamics at horizontal
scales between granulation and supergranulation, which have not been
covered by previous numerical simulations. A compressible 
fluid is used to provide a more realistic model of photospheric convection
than a Boussinesq fluid. Also, density stratification 
should attenuate the effect of an artificial bottom boundary
\citep{nordlund94}.

In Sect.~\ref{numerics} we present our numerical setup and physical
model. Section~\ref{results} is devoted to the analysis of the flow. 
The main consequences of the results are discussed in
Sect.~\ref{discuss}, which is followed by a short conclusion.


\section{\label{numerics} Numerical model and run parameters}
For the purpose of our investigations we use a
code designed to solve the fully compressible hydrodynamic equations
for a perfect gas \cite[\textit{e.~g.~}][]{cattaneo91} in cartesian geometry.
Constant dynamical viscosity and thermal conductivity are assumed.
A constant thermal flux is imposed at the bottom, 
while temperature is held fixed at the surface. The velocity field
satisfies stress-free impenetrable boundary conditions.
The initial state is a $m=1$ polytropic atmosphere ($\gamma=5/3$) 
with small random velocity $\vec{v}$, temperature $\theta$, and
density $\rho$ perturbations. The initial density contrast between the bottom and top plates
is 3, a value for which most of the 
features of stratification are already present in the linear 
convective instability problem \citep{gough76}.
The Prandtl number is $\pr=0.3$ and the Rayleigh number
evaluated in the middle of the layer is $\ray=3~10^5$
(650 times supercritical). We do not paste several smaller boxes
initially, as was the case in the paper by \cite{cattaneo01}, thus there is
no artificial spatial symmetry at $t=0$.

A sixth-order compact finite difference scheme \citep{lele92} is used in the
vertical (gravity $\vec{g}$) direction and a spectral scheme in the horizontal
(periodic) directions. FFTs are implemented  
\textit{via} the MPI version of FFTW  \citep{FFTW98}. 
Dealiasing by removal is performed using the 2/3 rule \citep{canuto}.
Time-stepping is done with a third-order, low-storage, fully
explicit Runge-Kutta scheme. Energy dissipation is handled
by laplacian terms, without any subgrid-scale
modelling or hyperviscosity. A very large aspect ratio $\aspra=42.6$ was
achieved using $82\times 1024\times 1024$ grid points. 
The simulation ran on 64 processors and a total of 400 GB of raw data was
collected throughout the numerical experiment.

\section{\label{results}Results}
In the following, the depth of the layer $d$ is used as unit of length
and the vertical thermal diffusion time $d^2/\kappa_{\mathrm{bot}}$
as unit of time ($\kappa_{\mathrm{bot}}$ is the
thermal diffusivity at the bottom of the layer).
\subsection{Flow structure and evolution}
\begin{figure}
\resizebox{\hsize}{!}{%
\includegraphics[width=5.cm]{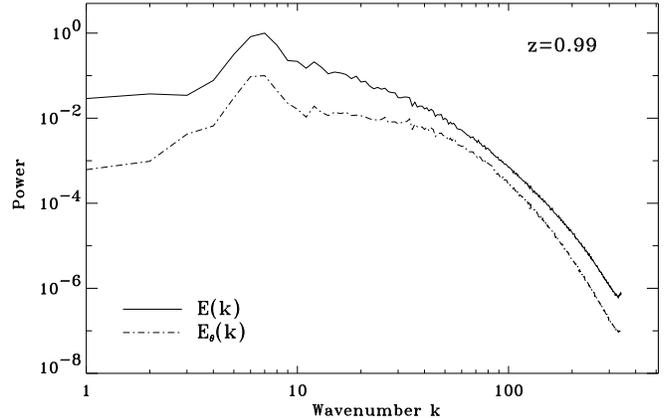}}
\caption{Normalized momentum $E(k,z)$ and temperature
  $E_\theta(k,z)$ power spectra at $t=0.7$ and  $z=0.99$ 
 (the temperature spectrum has been shifted). 
 The maximum around $k\sim 7$  corresponds to mesoscales. 
Also, a significant power excess in $E_\theta(k,z)$ is observed around
$k\sim 40$ close to the surface, in comparison to the deeper
layers. This feature is associated with granulation in this simulation.}
\label{spectre}
\end{figure}

\begin{figure}
  \resizebox{\hsize}{!}{%
\includegraphics[width=5.cm]{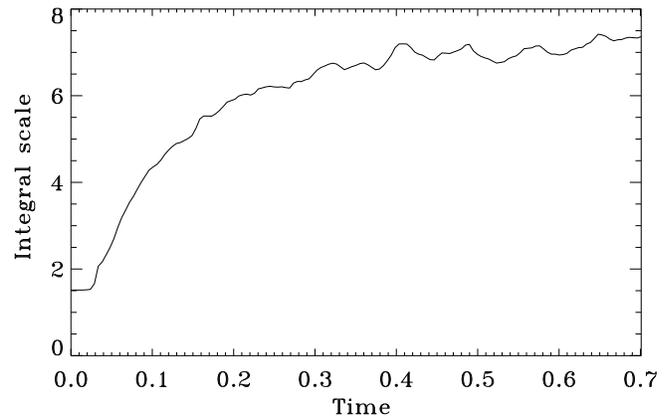}}
\caption{Integral scale $L_{\mathrm{int}}(z)\!=\!\frac{\pi}{2}\int
    k^{-1}E(k,z)\,\mbox{d}k\,\left/\,\int E(k,z)\,\mbox{d}k\right.$
    evolution at \hbox{$z=0.87$}. This scale represents the most energetic flow structures.} 
\label{intscale}
\end{figure}

\begin{figure*}
\centerline{\hbox{\includegraphics[width=6.2cm]{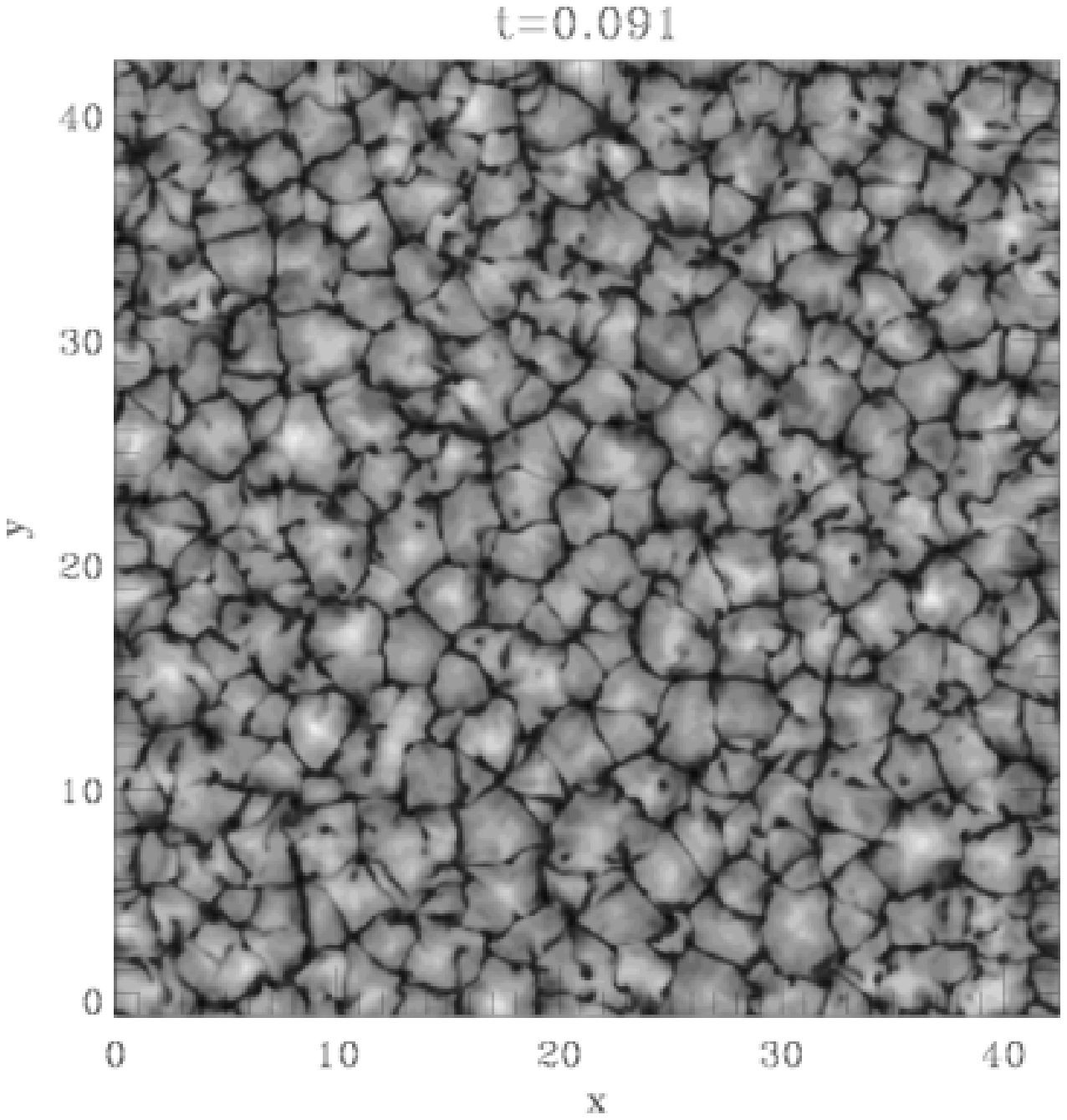}\hbox to -5pt{}
\includegraphics[width=6.2cm]{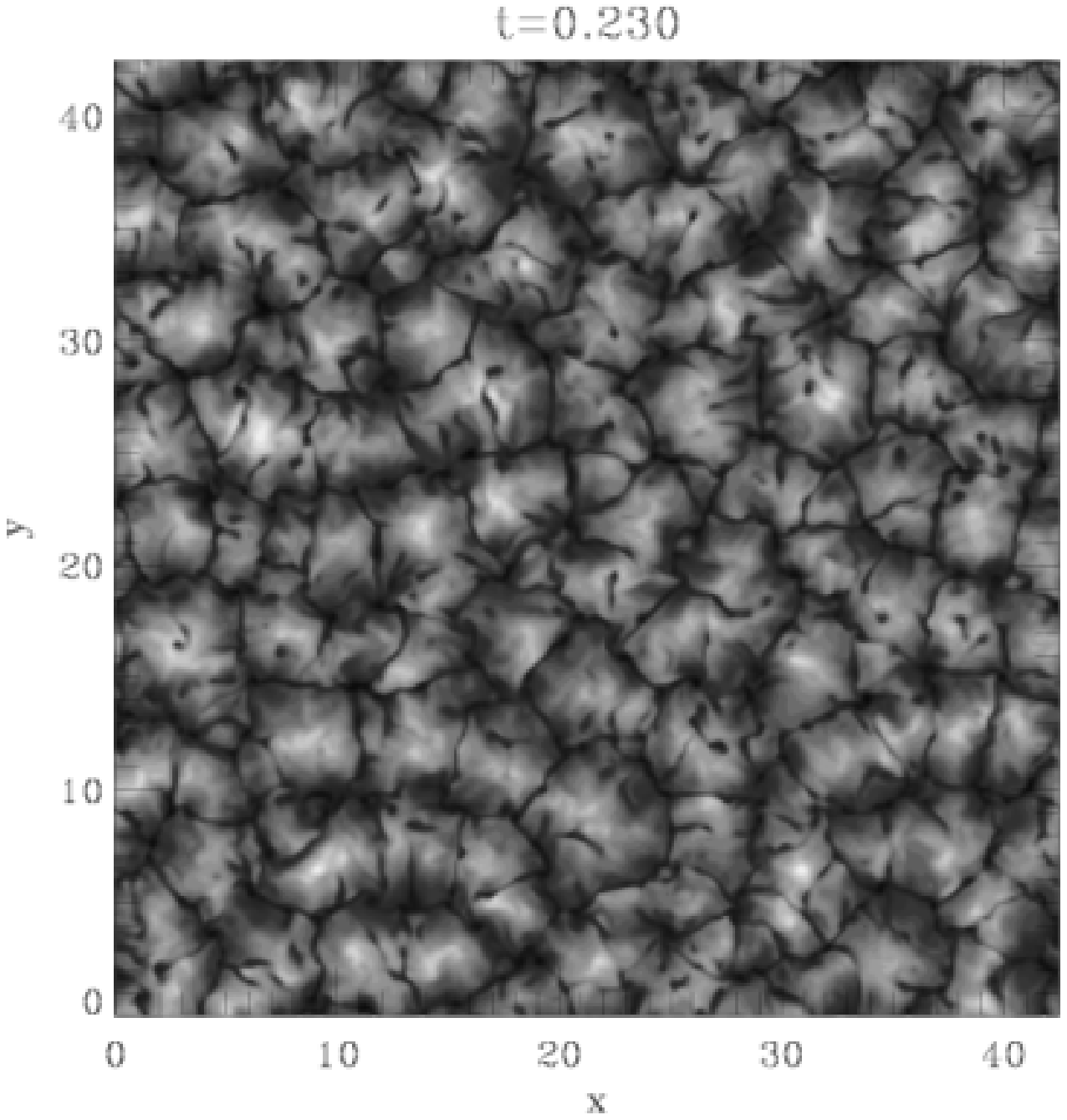}}}\vspace{-4pt}
\centerline{\hbox{\includegraphics[width=6.2cm]{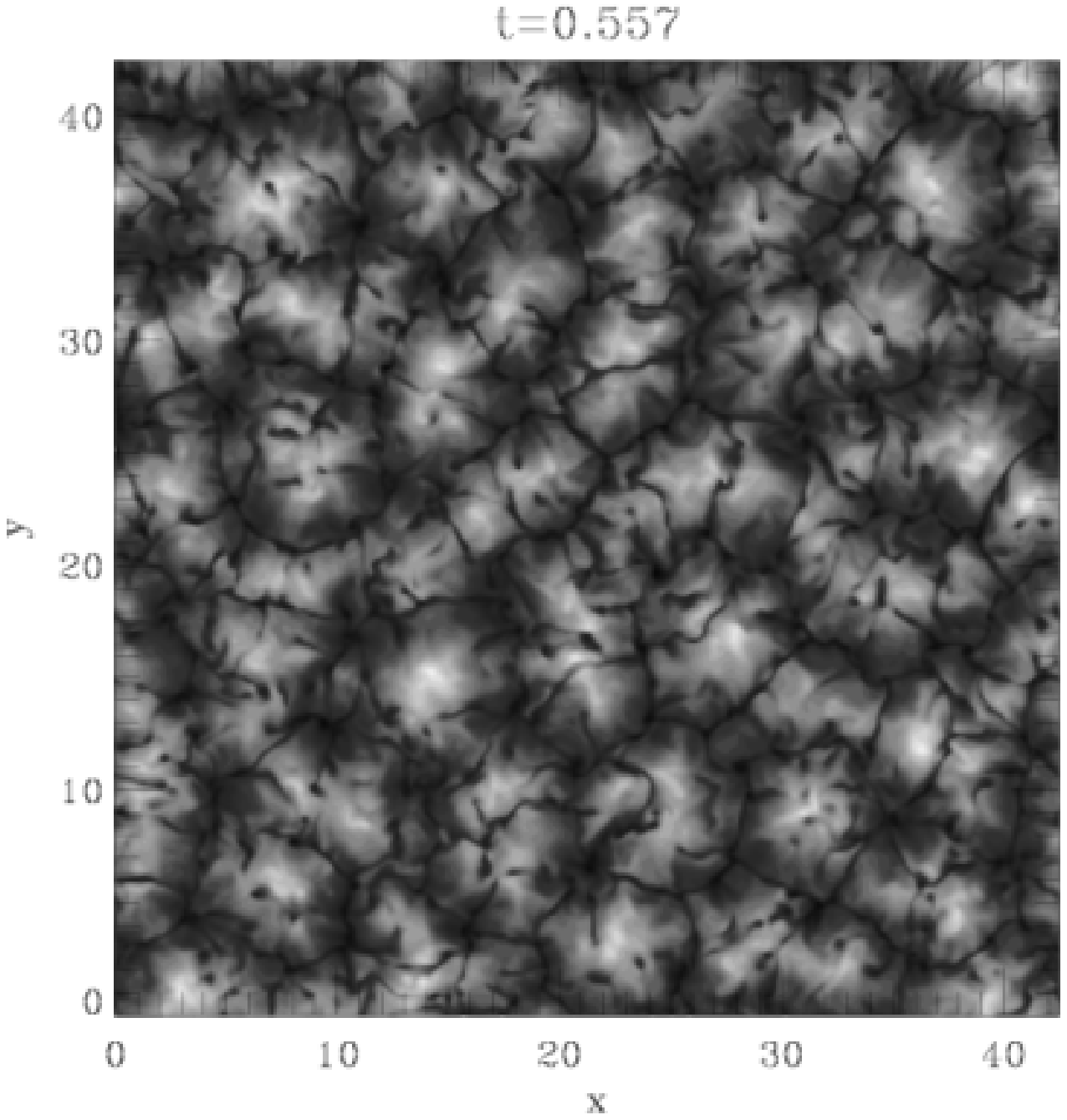}\hbox to -5pt{}
\includegraphics[width=6.2cm]{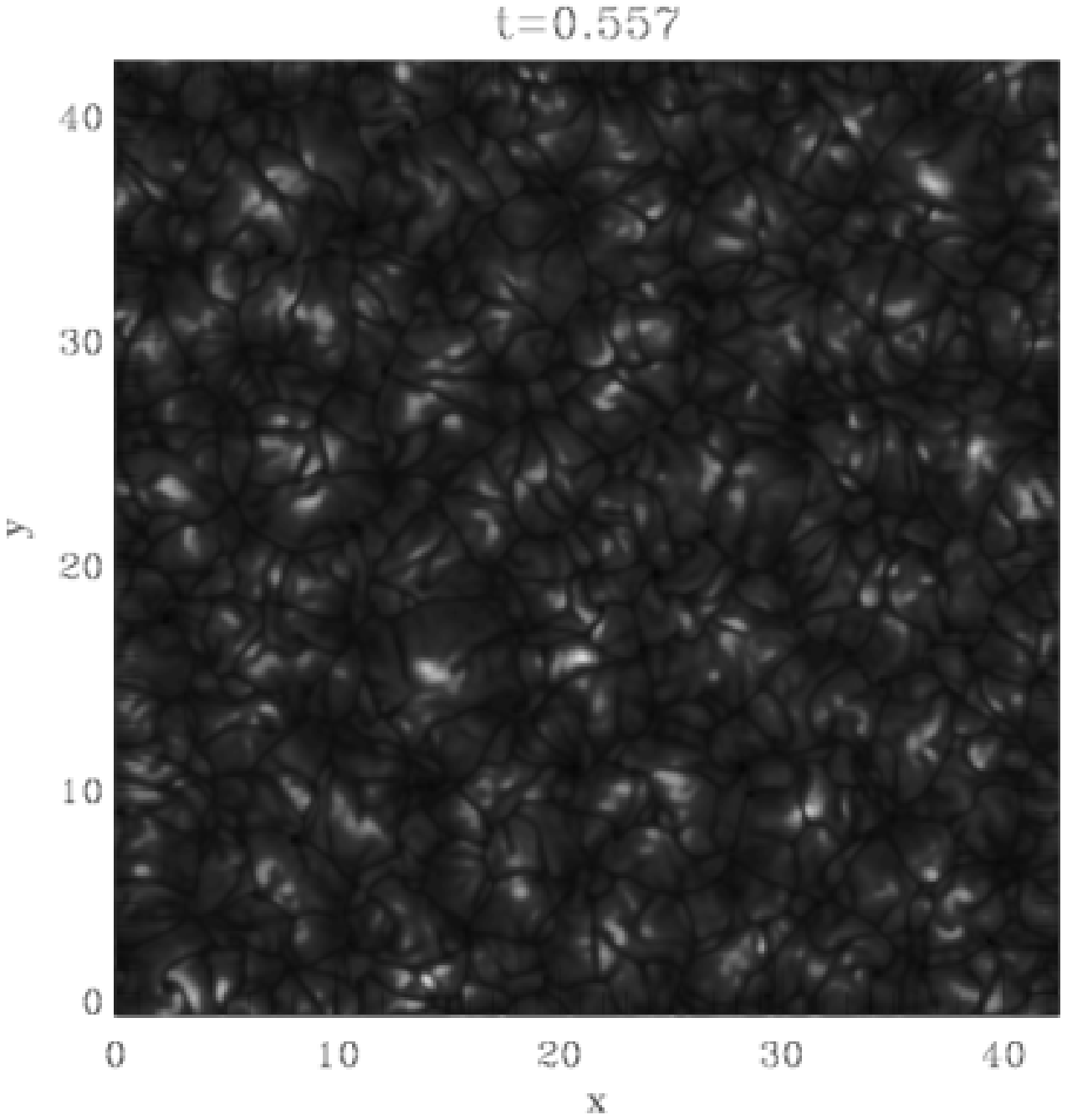}}}
\vspace{-4pt}
\caption{From left to right and top to bottom, time-evolution of
  temperature maps at $z=0.63$. The size of the visible
  mesoscale pattern increases until a quasi-steady state is reached (bottom-left image). 
  Bottom-right picture: same as bottom-left image, for a surface
  layer ($z=0.99$), showing the differences between in-depth
  mesoscale dynamics and the smaller-scale flow in the upper thermal boundary layer.}
\label{mesostruct}
\end{figure*}

We first describe the flow evolution during 0.7 thermal diffusion time,
corresponding to twelve turnover times (twice the vertical
crossing time based on the r.m.s. velocity). Initially, linear growth is observed
for a normalized horizontal wavenumber $k\sim 45$ (length $\ell\sim 1$) predicted by linear
theory. The maximum of the depth-dependent momentum spectrum (hats denote two-dimensional horizontal Fourier transforms)
\begin{equation}
\label{defspectre}
E(k,z)=\int_{\Omega_{\vec{k}}}\left|\widehat{(\rho\vec{v})}^{\phantom{*}}\!\!\null_{\vec{k}}(z)\right|^2k\,
\mbox{d}\Omega_{\vec{k}}
\end{equation}
then progressively shifts from
$k\sim 45$ to  $k\sim 5-8$ ($\ell\sim 5-8$)  at the end of the
simulation (Fig.~\ref{spectre}). 
As can be seen on Fig.~\ref{intscale}, the integral scale reaches
$L_{\mathrm{int}}\sim 7$.
Horizontal temperature maps at different times (Fig.~\ref{mesostruct})
clearly show that the dominant scale of the flow increases during the
simulation. This coherent pattern is best seen in
the middle of the layer, but most of the in-depth dynamics are 
still clearly visible in the surface layers. In the upper layers,
a second distinct smaller scale appears (bottom-right picture of
Fig.~\ref{mesostruct}). Its vertical extent (\hbox{0.2 $d$}) corresponds
to the surface region of superadiabatic stratification, while
the interior is almost isentropic.
Referring to the Sun, we shall identify this thermal boundary layer scale
with granulation, thus the larger internal scale is a
mesoscale.

Next we shall try to explain the growth and saturation of the
integral scale as well as the origin of the observed mesoscales.

\subsection{Convective origin of mesoscales}
\begin{figure}[h!]
\resizebox{\hsize}{!}{%
\includegraphics[width=5.cm]{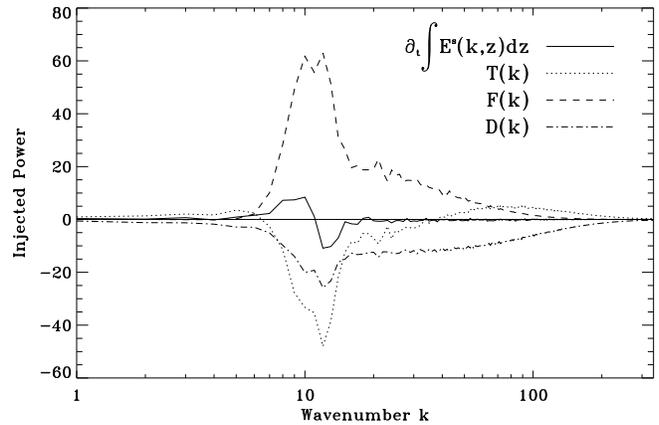}}
  \caption{Depth-integrated spectral transfer $T(k)$, buoyancy forcing
    $F(k)$, dissipation $D(k)$ and net power injected in $\int E^s(k,z)\,\mbox{d}z$
     following Eq.~(\ref{eqevol}). An average between $t=0.14$ and
     $t=0.20$ has been taken to outline the mean growth of $k<12$
     modes during this period owing to $F(k)$ (note especially that
     $T(k)<0$ for these scales). The maximum over $k$ of the 
     depth-integrated spectrum has been used for normalizing.}
  \label{transfer}
\end{figure}

After the linear growth, saturation of the velocity
amplitude at $k\sim 45$ occurs. At that time, this mode is 
the scale of energy injection. Larger scales are also highly linearly
unstable for \hbox{$\ray=3~10^5$} and we observe
that they continue to grow.
However, in the nonlinear regime, energy transfer 
to smaller scales limits this growth. 
To show this, we take the horizontal Fourier transform
of the momentum equation and extract its solenoidal
part by applying the projection operator
$\displaystyle{\mathcal{P}\left[\vec{v}\right]\,=\,\vec{v}-\grad\Delta^{-1}\div{\vec{v}}\,=\,\vec{v}^s}$,
denoted by $\mathcal{\widehat{P}}$ when acting on horizontally 
Fourier-transformed fields.
Taking the dot product with the complex conjugate of the solenoidal part
of momentum $\widehat{(\rho\vec{v})^s}^*\!\!\!\null_{\vec{k}}(z)$,
integrating over depth $z$ and
angles $\Omega_{\vec{k}}$ in the horizontal spectral plane, we obtain
a time-evolution equation for the solenoidal part $E^s(k,z)$ of $E(k,z)$, integrated vertically:
\begin{equation}
  \label{eqevol}
  \partial_t\int_0^1\!\!\!E^s(k,z)\,\mbox{d}z =  T(k)+F(k)+D(k),
\end{equation}
\begin{equation}
  T(k) =  \displaystyle{-2\,\re\left[\int_0^1\!\!\!\int_{\Omega_{\vec{k}}}\!\!\!\widehat{(\rho\vec{v})^s}^*\!\!\!\null_{\vec{k}}\cdot\widehat{\mathcal{P}}\left[\widehat{\div{\left(\rho\vec{v}\vec{v}\right)}}_{\vec{k}}\right]
    k\,\mbox{d}\Omega_{\vec{k}}\,\mbox{d}z\right]},
\end{equation}
\begin{equation}
 F(k) = \displaystyle{2\,\re\left[\int_0^1\!\!\!\int_{\Omega_{\vec{k}}}\!\!\!\widehat{(\rho\vec{v})^s}^*\!\!\!\null_{\vec{k}}\cdot\widehat{\mathcal{P}}\left[\widehat{\rho\vec{g}}_{\vec{k}}\right] k\,\mbox{d}\Omega_{\vec{k}}\,\mbox{d}z\right]}.
\end{equation}
$T(k)$ represents nonlinear transfers, $F(k)$ is the forcing by
buoyancy and $D(k)$, which has a similar definition, 
represents viscous dissipation. In Fig.~\ref{transfer}, we plot these 
quantities averaged over a time interval during which $k>12$ modes are steady, 
while modes with smaller $k$ develop. 
We observe that $F(k)$ is the basic energy supply on
large scales, as in the linear convective instability
mechanism. Nonlinear transfer $T(k)$ is always negative for 
modes with $7\leq k\leq 11$, which have a small but positive net energy growth. 
Therefore large scales do not come out of nonlinear interactions 
but have a \textit{convective} origin. This effect could also be observed 
using the energy equation: as in linear convection,  
energy is injected in large scales \textit{via} the 
advection of the horizontally averaged entropy profile, while
nonlinear transfers and diffusion only remove energy from them.

Note finally that the mesoscale pattern of Fig.~\ref{mesostruct}  is
expected to expand slightly  on a much longer time scale that can not
be achieved numerically. Also, the dominant scales may
depend on the Rayleigh number, as they result from a balance between
buoyancy and nonlinear transfers. 
These mesocells are very probably the same as
those observed by \cite{cattaneo01} in Boussinesq
simulations with $\aspra=20$. Their size is comparable in both
experiments. We therefore confirm these results for a compressible
fluid, in a larger aspect ratio box with no initial symmetry, but 
interpret them quite differently.
\section{\label{discuss}Discussion}
\subsection{Relations with solar photospheric convection}
We now discuss the relevance of our results to the Sun.
First, an estimate of the size (in km) of the typical 
structures of our simulation can be computed to clarify the comparison with
observations. Identifying the thermal boundary layer
thickness with the typical vertical extent of solar 
granulation ($\sim 150$ km), we find our mesocells to be 6~000 km wide. 

This numerical experiment represents a highly idealized model of photospheric 
convection, even though it integrates density stratification. Differences
with observations or with more realistic
simulations \citep{nordlund98} are therefore clearly expected. 
The most important one is the prominent peak at mesoscales in our power spectrum
(Fig.~\ref{spectre}), which might be due to boundary 
condition effects or to the absence of radiative transfer in our 
simulations. As noted by \cite{cattaneo01}, granulation is 
directly related to the formation of a thermal boundary layer, 
so that changes in boundary conditions might have a strong 
impact on the contrast between mesogranular and granular flows. Also, 
in radiative convection simulations by \cite{nordlund00}
with an open bottom boundary, the dominant scale visually increases
continuously with depth, which does not happen in our experiment.
Besides, the absence of radiative transfer in our simulations makes 
it impossible to define a $\tau=1$ surface. Actually, the intensity
map presented  by \cite{rieutord02}  does not exhibit clear mesoscale 
intensity modulation whereas temperature maps at fixed 
depth do. Since the $\tau=1$ surface does not correspond to a 
fixed depth,  mesoscale convective flows may be
present in the subsurface layers and be partly hidden this way.

Besides, various observations suggest some mesoscale organization.
\cite{oda84} has reported some clustering of granules around brighter
granules distributed on mesoscales.
In our simulation (Fig.~\ref{mesostruct}), the imprint of mesoscales 
is very clear at the surface and clustering of smaller granules also
occurs around bright spots corresponding to mesoscale upflows. 
Finally, a distribution of inter-network magnetic fields at the
scale of mesogranulation has been found recently  
\citep{dominguez2003,roudier2004}, which might also be related to strong 
convective mesoscale plumes. 

\subsection{Very large scales}
A final word must be said about very large-scale flows. According to
the previous estimates, the horizontal size of our box is
35~000 km, which does not leave much room for supergranules. 
We do not obtain a second peak at small $k$ in the spectra, 
so there is strictly speaking no supergranulation in our
simulation. This may be because we do not have the necessary
physical ingredients in the model, because our box is not 
large or deep enough, or because our run is too short. However,
noticeable positive nonlinear transfer occurs at the smallest $k$
(Fig.~\ref{transfer}) and we observed weak horizontally divergent 
large-scale flows of still unclear origin. When subjected to horizontal
strains, strong mesoscale vertical vortices resulting from 
angular momentum conservation in sinking plumes
\citep{toomre90} might help creating horizontal 
vorticity on large scales. 

\section{Conclusions}
The results of our simulations of large-scale convection in a
compressible polytropic atmosphere show that the dominant 
\textit{convective} mode is found at mesoscales. We observe that large 
aspect ratio simulations are necessary to study the convective
dynamics of these structures, since the integral scale is
$L_{\mathrm{int}}\sim 7$ in the final quasi-steady  state.
A slow evolution is expected on longer time
scales, which are unfortunately numerically out of reach.
Some similarities with solar observations
are found. If this kind of model is relevant to study the
solar photosphere, our results suggest that mesoscale convection may
be powerful below the Sun's surface. This  would help to 
explain the coexistence of two apparently distinct granular and
mesogranular scales. Supergranulation is not found in this experiment.




\begin{acknowledgements}
Numerical simulations have been carried out 
at IDRIS (Paris) and CINES (Montpellier). Both 
computing centers are gratefully acknowledged. 
F. R. would like to thank M. R. E. Proctor, 
N. O. Weiss, A. A. Schekochihin and B. 
Freytag for several helpful discussions.
\end{acknowledgements}

\bibliographystyle{aa}
\bibliography{lettre}

\end{document}